\documentclass[aps,pra,twocolumn,nofootinbib]{revtex4}
\usepackage{amsmath, amsthm, amscd, amssymb}
\usepackage{bm}
\usepackage{bbm}
\usepackage{epsfig}

\begin{document}

\title{Testing Vector Gravity with Gravitational Wave Interferometers}
\author{Anatoly A. Svidzinsky}
\affiliation{Department of Physics \& Astronomy, Texas A\&M University, College Station,
TX 77843}

\date{\today }

\begin{abstract}
Recently, the author proposed an alternative vector theory of gravity. To
the best of our knowledge, vector gravity also passes available tests of
gravity, and, in addition, predicts the correct value of the cosmological
constant without free parameters. It is important to find a new feasible
test which can distinguish between vector gravity and general relativity and
determine whether gravity has a vector or a tensor origin. Here we propose
such an experiment based on measurement of propagation direction of
gravitational waves relative to the perpendicular arms of a laser
interferometer. We show that transverse gravitational wave in vector gravity
produces no signal when it propagates in the direction perpendicular to the
interferometer plane or along one of the arms. In contrast, general
relativistic wave yields no signal when it propagates parallel to the
interferometer plane at $45^{\circ }$ angle relative to an interferometer
arm. The test can be performed in the nearest years in a joint run of the
two LIGO and one Virgo interferometers.
\end{abstract}

\maketitle

\section{Introduction}

Recently, the author proposed a new alternative vector theory of gravity
which assumes that Universe has fixed background geometry - four dimensional
Euclidean space, and gravity is a vector field in this space which breaks
the Euclidean symmetry \cite{Svid16}. Direction of the vector gravitational
field gives the time coordinate, while perpendicular directions are spatial
coordinates. Similarly to general relativity, the theory postulates that
gravitational field is coupled to matter universally and minimally through a
metric tensor which, however, is not an independent variable but rather a
functional of the vector gravitational field.

Such assumptions yield a unique theory of gravity. To the best of our
knowledge, it passes all available tests, including the recent detection of
gravitational waves by LIGO \cite{Abbo16,Abbo16a}. Vector gravity predicts
lack of black holes and gravitational waveforms measured by LIGO are
interpreted as being produced by orbital inspiral of massive neutron stars
that can exist in the theory. The measured waveforms can be fitted in the
framework of vector gravity as accurately as in general relativity \cite%
{Svid16}.

For cosmology vector gravity predicts the same evolution of the Universe as
general relativity with cosmological constant and zero spatial curvature.
However, vector gravity provides explanation of the dark energy as energy of
longitudinal gravitational field induced by the Universe expansion and
yields, with no free parameters, the value of $\Omega _{\Lambda }=2/3\approx
0.67$ which agrees with the recent Planck result $\Omega _{\Lambda
}=0.686\pm 0.02$ \cite{Planck14}. Thus, vector gravity solves the dark
energy problem.

Physical explanation of the dark energy in vector gravity is the following.
Expansion of the Universe yields change of spatial scale with time which can
be viewed as an increase of the distance between masses. This generates
matter current directed away from an observer. Such current induces
longitudinal vector gravitational field in a similar way as electric current
creates vector potential in classical electrodynamics. Average energy of the
longitudinal gravitational field induced by the Universe expansion is the
mysterious dark energy. Contrary to matter, it has negative energy density
and accelerates expansion of the Universe.

Despite of fundamental differences, vector gravity and general relativity
yield for the experimentally tested regimes quantitatively very close
predictions which allowed both theories to pass available tests. In
particular, vector gravity and general relativity are equivalent in the
post-Newtonian limit. At strong field, however, vector gravity substantially
deviates from general relativity and yields no black holes. Namely, the end
point of a gravitational collapse is not a point singularity but rather a
stable star with a reduced mass. One should mention that black holes have
never been observed directly and \textquotedblleft
evidences\textquotedblright\ of their existence are based on the presumption
that general relativity describes gravity for strong field. Until signatures
of the event horizons are found the existence of black holes will not be
proven.

Prediction of the matter behavior at strong gravity and interpretation of
the universe evolution on large scales depends on the theory of gravity we
are using. Thus, there is a need for a feasible test which can distinguish
between vector gravity and general relativity and rule out one of the two
theories. Here we propose such an experiment that can be done in the nearest
years using gravitational wave interferometers.

Both in general relativity and vector gravity the polarization of
gravitational waves emitted by orbiting binary objects is transverse, that
is wave yields motion of test particles in the plane perpendicular to the
direction of wave propagation. However, as we show, dependence of the laser
interferometer signal on the orientation of the interferometer arms relative
to the propagation direction of the gravitational wave is different in the
two theories.

Gravitational wave produces motion of the interferometer mirrors and changes
phase velocity of light. Both of these effects contribute to the relative
phase shift of light traveling in the perpendicular arms of the Michelson
interferometer. For certain propagation directions of the gravitational wave
relative to the arms the two contributions cancel each other yielding zero
net phase shift. Those are the directions of zero response of the
interferometer for which gravitational wave can not be detected for any
transverse polarization. As we show, directions of the zero response are
different for transverse gravitational waves in vector gravity and general
relativity. Detection of a wave in the direction of the zero response
predicted by a theory of gravity will rule out such theory.

\section{Response of interferometer on gravitational wave in vector gravity and general relativity}

In vector gravity for a weak transverse plane gravitational wave propagating
along the $x-$axis the equivalent metric reads \cite{Svid16}%
\begin{equation}
g_{ik}=\eta _{ik}+\left( 
\begin{array}{cccc}
0 & 0 & h_{0y}(t,x) & h_{0z}(t,x) \\ 
0 & 0 & 0 & 0 \\ 
h_{0y}(t,x) & 0 & 0 & 0 \\ 
h_{0z}(t,x) & 0 & 0 & 0%
\end{array}%
\right) ,  \label{met1}
\end{equation}%
where $\eta _{ik}$ is Minkowski metric and $h_{0y}$, $h_{0z}$ are small
perturbations obeying the wave equation. A rest particle (or mirrors of an
interferometer) will move under the influence of the gravitational wave (\ref%
{met1}) with a time-dependent velocity $V^{\alpha }=h_{0\alpha }c$ ($\alpha
=x$, $y$, $z$) perpendicular to the direction of the wave propagation. By
making a coordinate transformation into the co-moving frame of the test
particle 
\begin{equation*}
x^{\prime \alpha }=x^{\alpha }-\int^{t}V^{\alpha }dt,
\end{equation*}%
the metric (\ref{met1}) reduces to%
\begin{equation}
g_{ik}=\eta _{ik}+\left( 
\begin{array}{cccc}
0 & 0 & 0 & 0 \\ 
0 & 0 & h_{xy}(t,x) & h_{xz}(t,x) \\ 
0 & h_{xy}(t,x) & 0 & 0 \\ 
0 & h_{xz}(t,x) & 0 & 0%
\end{array}%
\right) ,  \label{met1a}
\end{equation}%
where $h_{xy}=h_{0y}$, $h_{xz}=h_{0z}$. Metric (\ref{met1a}) is written in
the coordinate system in which test particles do not move under the
influence of the gravitational wave.

On the other hand, in general relativity for a weak gravitational wave
propagating along the $x-$axis the metric in the co-moving frame evolves as 
\cite{Land95}%
\begin{equation}
g_{ik}=\eta _{ik}+\left( 
\begin{array}{cccc}
0 & 0 & 0 & 0 \\ 
0 & 0 & 0 & 0 \\ 
0 & 0 & h_{yy}(t,x) & h_{yz}(t,x) \\ 
0 & 0 & h_{yz}(t,x) & -h_{yy}(t,x)%
\end{array}%
\right) .  \label{wa3}
\end{equation}

\begin{figure}[t]
\begin{center}
\epsfig{figure=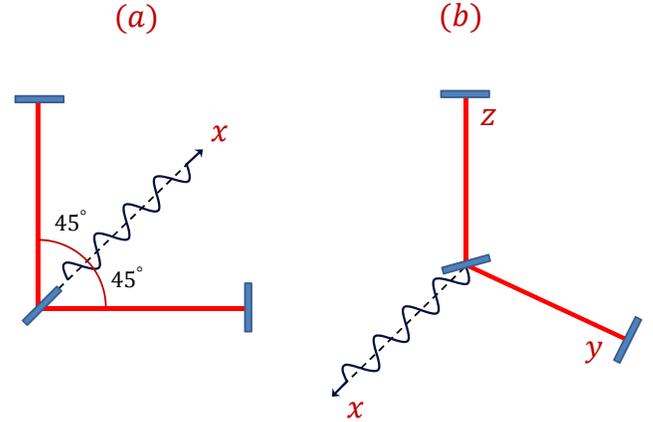, angle=270, width=9cm}
\end{center}
\par
\vspace{-1cm}
\caption{(a) Gravitational wave propagates in the interferometer plane at $%
45^{\circ } $ angle relative to the interferometer arms. Such wave produces
signal in vector gravity but not in general relativity. (b) Gravitational
wave propagates in the direction perpendicular to the interferometer plane.
Such wave yields signal in general relativity but not in vector gravity.}
\label{Arms}
\end{figure}

Here we investigate a response of a laser interferometer with perpendicular
arms on a gravitational wave in the two theories of gravity. In
gravitational field with a metric $g_{ik}$ the Maxwell's equations for the
electromagnetic field vector $A^{k}$ in the absence of charges read%
\begin{equation}
\frac{\partial }{\partial x^{k}}\left[ \sqrt{-g}g^{kl}g^{im}\left( \frac{%
\partial A_{m}}{\partial x^{l}}-\frac{\partial A_{l}}{\partial x^{m}}\right) %
\right] =0,  \label{max1}
\end{equation}%
while the Lorenz gauge equation is%
\begin{equation}
\frac{\partial }{\partial x^{k}}\left( \sqrt{-g}A^{k}\right) =0,
\label{max2}
\end{equation}%
where $A^{k}=g^{km}A_{m}$ and $g=$det$(g_{ik})$. Gravitational wave causes
oscillation of $g_{ik}$ in space and time. However, since frequency of the
gravitational waves is much smaller then frequency of the electromagnetic
waves traveling in the interferometer one can disregard derivatives of the
metric in Eqs. (\ref{max1}) and (\ref{max2}). Then Eqs. (\ref{max1}) and (%
\ref{max2}) reduce to%
\begin{equation}
g^{kl}\frac{\partial ^{2}A^{i}}{\partial x^{k}\partial x^{l}}-g^{im}\frac{%
\partial ^{2}A^{k}}{\partial x^{m}\partial x^{k}}=0,  \label{max3}
\end{equation}%
\begin{equation}
\frac{\partial A^{k}}{\partial x^{k}}=0.  \label{max4}
\end{equation}%
Combining them together we obtain%
\begin{equation}
g^{kl}\frac{\partial ^{2}A^{i}}{\partial x^{k}\partial x^{l}}=0.
\label{max5}
\end{equation}

In the co-moving frame there is no motion of the interferometer mirrors and
the phase shift of light traveling along the two arms appears due to
difference in the light phase velocity. Substitute $g^{kl}=\eta ^{kl}+h^{kl}$
in Eq. (\ref{max5}), where $h^{kl}$ is a small perturbation that has only
spatial components $h^{\alpha \beta }$ ($\alpha $, $\beta =x$, $y$, $z$),
yields the following propagation equation for the electromagnetic wave%
\begin{equation}
\left( \frac{1}{c^{2}}\frac{\partial ^{2}}{\partial t^{2}}-\nabla
^{2}+h^{\alpha \beta }\frac{\partial ^{2}}{\partial x^{\alpha }\partial
x^{\beta }}\right) A^{i}=0.  \label{max7}
\end{equation}%
In this equation, $h^{\alpha \beta }$ can be approximately treated as
constants since they vary slowly as compared to the fast variation of $A^{i}$%
. Looking for solution of Eq. (\ref{max7}) in the form $A^{i}\propto
e^{-i\omega t+i\mathbf{k}\cdot \mathbf{r}}$, where $\mathbf{k}$ is the wave
vector of the electromagnetic wave, we obtain the following dispersion
relation for light%
\begin{equation*}
\frac{\omega ^{2}}{c^{2}}=k^{2}-h^{\alpha \beta }k_{\alpha }k_{\beta },
\end{equation*}%
and, hence, the phase velocity of the electromagnetic wave is (see also \cite%
{Gert63})%
\begin{equation}
V_{\text{ph}}=\frac{\omega }{k}\approx c\left( 1-\frac{1}{2}h^{\alpha \beta }%
\hat{k}_{\alpha }\hat{k}_{\beta }\right) ,  \label{max8}
\end{equation}%
where $\hat{k}=\mathbf{k}/k$.

Equation (\ref{max8}) shows that presence of the gravitational wave leads to
the change of the phase velocity of light which depends on the direction of
the light propagation $\hat{k}$. If arms of the interferometer are oriented
along unit vectors $\hat{a}$ and $\hat{b}$ then difference in the phase
velocities of the laser light propagating along the two arms is 
\begin{equation*}
\Delta V_{\text{ph}}=\frac{c}{2}h^{\alpha \beta }\left( \hat{a}_{\alpha }%
\hat{a}_{\beta }-\hat{b}_{\alpha }\hat{b}_{\beta }\right) .
\end{equation*}

Signal of the LIGO-like (Michelson) interferometer with arms of length $L$
oriented along the directions $\hat{a}$ and $\hat{b}$ is proportional to the
relative phase shift $\Delta \varphi $ of electromagnetic waves traveling a
roundtrip distance $2L$ along the two arms \cite{Gert63} 
\begin{equation}
\Delta \varphi =k\frac{2L}{c}\Delta V_{\text{ph}}=\frac{\omega L}{c}%
h^{\alpha \beta }\left( \hat{a}_{\alpha }\hat{a}_{\beta }-\hat{b}_{\alpha }%
\hat{b}_{\beta }\right) ,  \label{wa4}
\end{equation}%
where $\omega $ is the frequency of electromagnetic wave and $h^{\alpha
\beta }$ is the spatial perturbation of the metric in the reference frame in
which interferometer mirrors do not move (frame of Eqs. (\ref{met1a}) and (%
\ref{wa3})).

For the gravitational wave propagating along the $x-$axis, Eq. (\ref{wa4})
yields for the gravitational wave (\ref{wa3}) in general relativity%
\begin{equation}
\Delta \varphi =\frac{\omega L}{c}\left[ h^{yy}\left( \hat{a}_{y}^{2}-\hat{b}%
_{y}^{2}+\hat{b}_{z}^{2}-\hat{a}_{z}^{2}\right) +2h^{yz}\left( \hat{a}_{y}%
\hat{a}_{z}-\hat{b}_{y}\hat{b}_{z}\right) \right] ,  \label{wa5}
\end{equation}%
while for the transverse wave (\ref{met1a}) in vector gravity we obtain%
\begin{equation}
\Delta \varphi =\frac{2\omega L}{c}\left[ h^{xy}\left( \hat{a}_{x}\hat{a}%
_{y}-\hat{b}_{x}\hat{b}_{y}\right) +h^{xz}\left( \hat{a}_{x}\hat{a}_{z}-\hat{%
b}_{x}\hat{b}_{z}\right) \right] .  \label{wa6}
\end{equation}

Equations (\ref{wa5}) and (\ref{wa6}) show that vector gravity and general
relativity predict qualitatively different effect of the gravitational wave
on the interferometer signal. Namely, general relativistic gravitational
wave of any polarization (arbitrary $h^{yy}$ and $h^{yz}$) produces no
signal when gravitational wave propagates parallel to the interferometer
plane at $45^{\circ }$ angle relative to one of its perpendicular arms (see
Fig. \ref{Arms}\textit{a}). E.g., this is the case for $\hat{a}=\left(
1,1,0\right) /\sqrt{2}$ and $\hat{b}=\pm \left( 1,-1,0\right) /\sqrt{2}$.
For these orientations the gravitational wave in vector gravity will produce
signal, namely, 
\begin{equation}
\Delta \varphi =\frac{2\omega L}{c}h^{xy}.
\end{equation}

On the other hand, gravitational wave in vector gravity (for arbitrary $%
h^{xy}$ and $h^{xz}$) yields no signal if gravitational wave propagates in
the direction perpendicular to the interferometer plane, e.g. $\hat{a}%
=\left( 0,1,0\right) $ and $\hat{b}=\left( 0,0,1\right) $ (see Fig. \ref%
{Arms}\textit{b}), or along one of the interferometer arms, e.g. $\hat{a}%
=\left( 1,0,0\right) $, $\hat{b}=\left( 0,0,1\right) $ or $\hat{a}=\left(
0,1,0\right) $, $\hat{b}=\left( 1,0,0\right) $. For these orientations the
gravitational wave in general relativity can produce signal.

\begin{figure}[t]
\begin{center}
\epsfig{figure=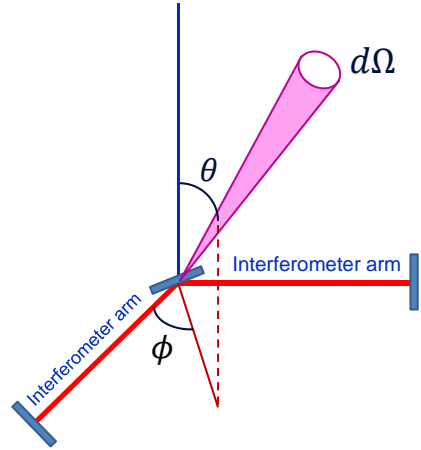, angle=270, width=9cm}
\end{center}
\par
\vspace{-1cm}
\caption{Spherical coordinate system in the frame of interferometer arms.}
\label{N}
\end{figure}

This difference can be used to test theories of gravity in future
polarization experiments with several LIGO-like interferometers. The
experiment can be conducted with three interferometers. Simultaneous
detection of the gravitational wave by all three instruments allows us to
determine the direction of the wave propagation by measuring the wave
arrival times at the interferometer locations. By detecting many of such
events one can collect statistics and find a distribution of the direction
of the detected gravitational waves relative to the interferometer arms.
Namely, one can measure the distribution function $N(\theta ,\phi )$ defined
as $dN=$ $N(\theta ,\phi )d\Omega $, where $dN$ is the number of events for
which detected gravitational waves propagate inside the solid angle $d\Omega
=\sin (\theta )d\theta d\phi $. Here $\theta $ and $\phi $ are the polar and
azimuth angles in the spherical coordinate system in the frame of the
interferometer arms (see Fig. \ref{N}).

Vector gravity predicts that such distribution will have dips in the
directions perpendicular to the interferometer plane - $\theta =0$, $\pi $
(Fig. \ref{Arms}\textit{b}) and along the interferometer arms - $\theta =\pi
/2$, $\phi =0$, $\pi /2$, $\pi $, $3\pi /2$. The dips appear because for
these propagation directions the interferometer can not detect the
transverse gravitational wave. In the case of general relativity the dips
will appear in the directions for which wave propagates in the
interferometer plane at $45^{\circ }$ angle relative to one of the
interferometer arms ($\theta =\pi /2$ and $\phi =\pi /4$, $3\pi /4$, $5\pi
/4 $, $7\pi /4$). Thus, such experiment is able to distinguish between the
two theories.

One should mention that vector gravity also predicts existence of
longitudinal gravitational waves which, however, are not emitted by orbiting
binary starts. For such a wave propagating along the $x-$axis the equivalent
metric in the co-moving frame reads \cite{Svid16}%
\begin{equation}
g_{ik}=\eta _{ik}+\left( 
\begin{array}{cccc}
0 & 0 & 0 & 0 \\ 
0 & -2h(t,x) & 0 & 0 \\ 
0 & 0 & h(t,x) & 0 \\ 
0 & 0 & 0 & h(t,x)%
\end{array}%
\right) .
\end{equation}%
Longitudinal gravitational waves produce no interferometer signal if they
propagate at equal angles relative to the interferometer arms (more exactly
when $\hat{k}\cdot \hat{a}=\pm \hat{k}\cdot \hat{b}$). Thus, propagation
direction perpendicular to the interferometer plane shown in Fig. \ref{Arms}%
\textit{b} is the direction of zero response for all kind of gravitational
waves in vector gravity. However, waves propagating along the arms can
produce signal if they are longitudinal.

\section{Conclusion}

To conclude, even though in both theories the gravitational waves generated
by orbiting stars are transverse, vector gravity and general relativity
predict qualitatively different dependence of the gravitational wave signal
on the orientation of the laser interferometer arms. Gravitational wave
produces motion of the interferometer mirrors and changes phase velocity of
light. Both of these effects contribute to the interferometer signal. For
certain directions of the gravitational wave propagation the two
contributions cancel each other yielding zero net phase shift for any wave
polarization. These directions of the zero response are different in the two
theories.

We find that general relativistic gravitational wave produces no signal when
it propagates parallel to the interferometer plane at $45^{\circ }$ angle
relative to one of the arms (see Fig. \ref{Arms}$a$). On the other hand,
transverse gravitational wave in vector gravity yields no signal when it
propagates in the direction perpendicular to the interferometer plane (Fig. %
\ref{Arms}$b$) or along one of the interferometer arms. The former
propagation direction yields no signal even if the wave is longitudinal - it
is the direction of the zero response for all type of waves in vector
gravity. This difference can be used to distinguish between general
relativity and vector gravity in polarization experiments with gravitational
wave interferometers.

Such experiment is crucial for our understanding of the nature of gravity
and can test whether gravity has a tensor or a vector origin. Simultaneous
detection of gravitational waves in at least three instruments is necessary
for the experiment. A joint scientific run of the two LIGO interferometers
in the US and the Virgo interferometer in Italy is capable of distinguishing
between tensor and vector origin of gravity.

I thank Dennis R\"{a}tzel for valuable discussions and gratefully
acknowledge support of the Office of Naval Research (Award Nos.
N00014-16-1-2578 and N00014-16-1-3054) and the Robert A. Welch Foundation
(Award A-1261).

\end{document}